# A New Universal Constant Determining Expansion of the Universe

Charles B. Leffert, Emeritus Professor, Wayne State University
Contact: 1302 Wrenwood Drive, Troy, Michigan 48084

A new universal constant of expansion has been discovered with amazing predictive power once its density-time relations have been deciphered. The new constant is $\kappa = G\rho t^2$ where G is the gravitational constant, $\rho$ is the average total mass-energy density of our universe and t is cosmic time. With the ten parameters known, this relation promises to account for the expansion of our universe from its beginning into the far future. The most important and most difficult item is time and its scaling relation with the densities. The new cosmological theory will be presented in this paper to show good predictions of the cosmological parameters. The theory will be used in a second paper to show that acceleration of the expansion rate is not needed to account globally for the exploding-star supernova Ia radiation that has traveled such great distances in our expanding universe.

## 1.0 Introduction

### 1.1 Philosophical considerations of a: "First Law of Cosmology"

The concept of a "law" of cosmology has been questioned on a number of grounds. In a recent paper George F. R. Ellis [1] argues that a physical law gains acceptability by confirming repeated physical measurements. If there is only one universe, and we cannot measure its initial or boundary conditions, then there can be no confirmation of the claim of "law of cosmology". On the nature of "existence" he points out that the contents of our universe are not well understood at any time in its expansion. Also the local laws of physics, as a basis for cosmology, are questionable as is the predictions of infinite densities.

The theory to follow avoids these objections by the following: Assumption 1: Our spatially three-dimensional (3-D) universe was born in an altogether different, higher dimensional (m-D), background "epi-universe" by a dynamical process that continues today to expand our 3-D universe and whose expansion rate (and the m-D process) can certainly be measured. Assumption 2: Except for the distribution of condensed mass, another such 3-D universe could be expanding exactly like ours in this background epi-universe and will eventually merge with ours. Two universes can merge forming a common 3-D surface universe and, in principle, the blue-shifted radiation of approaching foreign stars could be measured. Merging of other young universes could re-supply portions of our universe with new hydrogen and, in principle, maintain life long after the present supply of hydrogen is exhausted.

### 1.2 The new theory of expansion

The late Dennis Sciama said, that we do not yet have a tight cosmological theory [2]. The proposal of this paper is that now we do.

A new cosmological theory has been developed for the beginning and expansion of our universe. It is a rigid theory with only one solution, so it is either correct or it is wrong. To date, comparisons to measurements show it to be correct. The story of this discovery [3, 4, 5] will be reviewed in two papers. In this first paper, the derivation of the theory, with discovery of the universal constant of expansion will be described and in the second paper the new theory will be extended to show that measurements of global phenomena also confirm its predictions.

The universal constant of expansion kappa, $\kappa$, is a dimensionless number,

$$\kappa = G\rho t^2, \tag{1}$$

where G is the gravitational constant, $= 6.673 \times 10^{-8}$ cm$^3$ g$^{-1}$ s$^{-2}$, $\rho$ is the summation of all of the mass-energy densities in our universe, and t is cosmic time. Equation (1) governs the expansion of the universe. It says, given the first of its nine parameters, the value of kappa $\kappa$ will be known and this relation will tell exactly how the 3-D universe expanded from its beginning to the present and on into the far future, that is, R(t) or t(R), where R is the scale factor for the expansion of the universe.

Kappa Listing: Parameters Needed to Use Eq. (1)

Contents of 3-D Universe
    1. Their number (eg., i = 3)
Scaling of densities with the expansion, $\rho_i \propto R^{-Ni}$
    2. i = 1
    3. i = 2
    4. i = 3
Values of densities, $\rho_{i0}$ (eg., at present)
    5. i = 1
    6. i = 2
    7. i = 3
Definition of square of cosmic time
    8. $t^2 \equiv$
Value of cosmic time (e.g., at present $t_0$)
    9. $t_0 =$
Value of universal constant
    10. $\kappa =$

Represented this way, in hindsight, satisfying the first nine statements would be a difficult task and not very encouraging. After a century of trying, theorists are still working to develop many cosmological models and there are still none with a fully satisfactory prediction of R(t) or t(R).

Notice that time appears in the universal constant and it appears squared. What time represents physically seems to be unknown. In current physics time is a symmetric mathematical parameter. The Friedmann equation [6] of general relativity (GR) expresses the Hubble parameter squared, $H^2$, where H has units of inverse time $t^{-1}$. Indeed, in the early twentieth century, attention was given to the dimensionless grouping $G\rho\tau^2$ where $\tau = H^{-1}$, especially by E. A. Milne [7] but it was not generally adopted.

This paper is arranged to show first in Section 2, the vision of how our universe came to be. Section 3 presents the derivation of the universal constant of expansion and the follow-on derivation of a new theory of the expansion or our universe. Section 4 presents some further details about cosmic time and some surprising features of the universal constant. Section 5 gives a geometric analysis for further theoretical support for the new theory. Section 6 explains why the present big bang theory, based on general relativity, is having such trouble predicting current measurements of supernova Ia radiation. Section 7 presents the Summary and Conclusions.

The new cosmological theory given in the following text, will be used in the second paper to see if it can correctly predict the supernova Ia data without the current problems of added unknown "dark energy" and without its present acceleration of the expansion rate of our universe.

## 2. Vision of birth of our 3-D universe

The following vision flowed from a long-held intuition that gravity was somehow due, not to an attractive force, but due to bombardment of particles from outside our 3-D universe.

If our 3-D universe had a beginning, there had to be something else existing before it did. There was an older, very different "epi-universe" already in existence of higher (m-D) spatial dimensions. At a certain epi-time, a spontaneous symmetry-breaking event occurred that produced a very small new foreign object (occupying less epi-space) that became a catalytic site for an exponential production on further such new sites. Using Planck's natural units of $l_p$ and $t_p$ from our 3-D space-time, call these new epi-objects 4-D spatial hypercubes of edge-length $l_p$ that produce another such hypercube every $t_p$ second. Very quickly (~$10^{-33}$ s) all of these "free" 4-D cells were forced into a 4-D ball of radius R, which continues to expand because the surface of the ball (our new 3-D universe of unknown 4-D ($nl_p$) thickness) is now the foreign object in the older epi-universe. This new epi-dynamic, called "Spatial Condensation (SC)" continues on its surface and on all (epi-foreign) mass contents within its 3-D surface.

This concept of a beginning had to be modified to include a random element in the form of two types of 4-D cells, an x-type and a c-type where the first 4-D cell produced was of the x-type that produced another x-type randomly but on average of one of four reproductions and otherwise produced c-type acceptable to the 4-D ball. The c-type of 4-D cell reproduces only more c-type 4-D cells. In this scenario, the 4-D ball accepts and grows with the addition of the c-type cells and the 4-D ball rejects as flotsam the growing x-type cells and the new mass contents of the 3-D surface. Figure 1 depicts the overall scheme. Our 3-D universe is still growing rapidly in the mother epi-universe.

Spatial condensation is a new dynamic producing *space* both outside and inside our 3-D universe and our 3-D universe is no longer isolated. Globally, many concepts will change including the concept of the conservation of 3-D energy. Present physics calls *space* a "vacuum" which is said to "stretch" without limit or source. Besides the characteristic time of our 3-D universe with its limiting velocity c, it can be expected that the epi-universe acts on an altogether faster epi-time with its limiting velocity $c^+ >> c$.

Fortunately, to use the universal constant and theory for the expansion of our 3-D universe, the details of the spatial condensation process are not needed. It is sufficient to know only the production rate versus time of its 4-D cells.

## 3. Building the new SC-theory and derivation of kappa

The vision requires a new 3-D scaling relation to account for the continued production of the 4-D x-stuff mass. The unknown "dark matter" of present physics is replaced with the 4-D x-stuff or "dark mass" that is within, but on the inner 4-D edge, of our universe and it is set to scale with the expansion as $\rho_x(R) \propto R^{-2}$. The density decreases, but the total mass increases. Since x-stuff reproduces only on x-stuff, this means it continues to grow as clumps that form excellent seeds for the condensation of ordinary matter and the production of very early 4-D black holes.

Our 3-D universe now has i = 3 kinds of content where radiation and ordinary matter scale as in the big bang (BB) theory as $\rho_r(R) \propto R^{-4}$ and $\rho_m(R) \propto R^{-3}$ and the new third content as $\rho_x(R) \propto R^{-2}$. Now comes the crucial part, our 3-D cosmic *time* t must be defined. At this point, there is no indication that a universal constant even exists.

First, the expansion of the universe is its own clock, the keeper of the asymmetric cosmic time. The change in the periodic motions of ordinary clocks and atoms with local changes in velocity produces negligible change of expansion on the cosmic scale. Next, it is noted that in Planck's natural units $G\rho_p t_p^2 = 1$. It seems that the *time* for anything that happens is inversely proportional to the resistance that keeps it from happening faster – be it even the filling of the bathtub with water by adjusting the resistance to flow at the faucet. We know from BB-theory that the early universe was dominated by radiation and much later by matter. But with the new scaling, the universe is now dominated by dark mass and will be on into the future. Equation (1) is not a wave equation so it certainly does not predict gravitational waves.

All of these contents of our universe are acting in parallel, probably with their own partial time and resistance to spatial condensation. As electrical resistors in parallel add by their inverse values, perhaps the cosmic partial times $\Gamma_i$ add by the inverse squares of their partial times to give the inverse square of cosmic time (never, cosmic time t = 0),

$$t^{-2} = \sum_i \Gamma_i^{-2} \qquad \Gamma_i^2 = (\kappa/G)/\rho_i(R), \qquad (2)$$

where $\rho_i$ are given by Eqs. (A.9) to (A.11) of Table A. Summation produces the universal constant of expansion of Eq. (1). Present values of the densities of radiation and matter are given in Table A but not that of x-stuff, $\rho_{x0}$.

One can learn much from this relation even if one does not know the numerical value of kappa but one must be careful of its relation. For example, from Eq. (1), ($\kappa = G\rho t^2$) = ($\kappa = G\rho_0 t_0^2$) or,

$$t = t_0(\rho_0/\rho(R))^{1/2}, \qquad (3)$$

the same as for the Einstein - de Sitter universe, except that $\rho$ now contains a new mass, $\rho_x$. After simple derivatives with respect to cosmic time one gets, explicitly in terms of cosmic time, the Hubble parameter (the over-dot represents d/dt),

$$H \equiv \dot{R}/R = (1/t)(\rho/\rho_2), \qquad \tfrac{1}{2} \leq Ht \leq 1 \quad \text{as } t \to \infty, H \to 0, \qquad (4)$$

where $\rho$ and $\rho_2$ are defined in Table A. In Table A, equations are also given for the dimensionless expansion rate $\dot{R}/c$ with a constant 3-D limit of $\dot{R} = c$.

A second time derivative of R in Eq. (4) produced Eq. (A.18) for the deceleration q of our universe where, $1 \geq q \geq 0$. In agreement with $\dot{R}/c$, the deceleration rate q has a very low present value of $q_0 = 0.0084$ and also predicts a zero future limit. **There is now, and into the future, no acceleration of the 3-D expansion rate.**

For comparison of Eq. (4) to the BB-Friedmann universes for q >1/2 to q < ½, then tH < 2/3 to tH > 2/3, and for deceleration q = ½, tH = 2/3. This last case represents an SC-intermediate matter-dominated universe and SC- Eqs. (A.16) and (A.18) agree.

In the present consensus BB-flat, infinite universe, expansion and radiation occur radially away from any 3-D point. In contrast for the closed 3-D universe, the source of the SC-expansion is in the fourth radial direction, normal to any motion in our 3-D universe. Therefore all radiation from a strong real source will be redshifted and focused as a virtual energy source at the antipode of the real source.

Without the value of kappa to complete the theory, it was difficult to fix the present density of x-stuff and other parameters. Then the value of $H_0$ ranged from 40 to 100 km s$^{-1}$ Mpc$^{-1}$ and the age $t_0$ from 10 to 20 Gy. Dark matter (now dark mass) was thought to be a factor of about 10 larger than the mass of ordinary matter. The value of kappa was needed.

In the book "Gravitation," the value of kappa $= 3/32\pi$ was found on page 733 as in the following quote:

> It is a striking feature of the radiation-dominated era of the early Friedmann universe that the density depends upon time according to a simple universal law,
> $$\rho_r = 3/32\pi t^2 \quad (27.53)$$
> (final line and final column of Box 27.3).
>
> From *Gravitation* BY Charles Meisner, Kip Thorne and John Wheeler, ©1973 by W.H. Freeman and Company. Used with permission.

The gravitational constant G does not appear here because these authors use units where G = 1. Of course, $3/32\pi$ was not recognized as a universal constant, because in their Box 27.3, the equivalent Friedmann density for matter is shown as $\rho_m = 1/6\pi t^2$.

In the SC-theory the new value of kappa was put to work to predict the current density of x-stuff or dark mass by difference,

$$\rho_0 = \kappa/Gt_0^2 \qquad \rho_{x0} = \rho_0 - \rho_{r0} - \rho_{m0}. \qquad (5)$$

A present value of the radiation density $\rho_{r0} = 9.40 \times 10^{-34}$ g cm$^{-3}$ had been obtained from the present temperature of the CMB radiation and a present value of the baryon matter density $\rho_{m0} = 2.72 \times 10^{-31}$ g cm$^{-3}$ had been obtained from reported early calculations of the nucleosynthesis era [8, 9]. Assuming our universe was already very near its constant rate of expansion $\dot{R} = c$, so that the new dark mass was dominant, the present size of our universe was approximated from Eq. (4), $R_0 = ct_0(\dot{R}/c)_0(\rho_2/\rho)_0$. The last two factors at present, are each slightly greater than unity, with limits of unity, so replace both with $(\rho_{T0}/\rho_{x0})^{1/2}$, which also has a limiting value of unity or,

$$R_0 \approx ct_0(\rho_0/\rho_{x0})^{1/2}, \qquad \text{then} \quad R = R_0/(1 + Z). \qquad (6)$$

Only a good value of the present age of our universe $t_0$ was needed now to complete the SC-theory. In the June, 1998 issue of Physics Today, a paper by Bertram Schwarzschild [10] showed the astronomical measurements of the radiation from 40 distant exploding stars of a class called supernova Ia. A plot was shown together with a number of theoretical curves representing various universes of the standard BB-model with varying range of density parameters ($\Omega_m$, $\Omega_\Lambda$) where $\Omega_i = \rho_i/\rho_c$ and $\rho_c = 3H^2/8\pi G$ and the subscript $\Lambda$ refers to chosen values of Einstein's constant, lambda.

Those measurements were of an effective magnitude $m_{eff}$ of the radiation versus redshift Z of the host galaxy. That radiation had traveled great distances through the expanding universe and was of just the type needed to confirm R(t) of this new theory. An early analysis of these SNIa reported data [3] showed a satisfactory SC-theoretical fit to the data set with the age of our 3-D universe at $t_0$ = 13.5 Gy [4]. This value of $t_0$, gave $\rho_{x0}$ = 2.19x10$^{-30}$ g cm$^{-3}$, $\rho_{T0}$ = 2.46x10$^{-30}$ g cm$^{-3}$, $H_0$ = 68.6 km s$^{-1}$ Mpc$^{-1}$ and total mass energy $E_{m0} = \rho_0 c^2$ = 1.09x10$^{77}$ ergs. These early predictions for our universe of mass density, age, and Hubble constant [3] were in good agreement with the latter Wilkinson Microwave Anisotropy Probe (WMAP) measurements [11]. Further development of the SC-theory to account for the 4-D tortuous path of the photons and analysis of later measurements will be made in paper 2.

Figure 2 is a striking log-log plot that shows the changing domination of our universe by its contents. For all of the figures from Fig. 2 to Fig 8, the curves representing the SC-expansion of our 3-D universe were obtained by solving Eq. (1) with the listed parameters fixed as in Table A and using the value of $\kappa$. No integration of a differential equation was necessary. We learn about our universe from radiation emitted in the distant past. Of course, we do not receive radiation from the future but that very radiation we do measure continues on into the future and could also be measured again by hypothetical astronomers of the future. Thus we can, and should for consistency, project our model calculations into the future, when all of the parameters have changed, and see if the theory predicts for the past what has already been measured.

For the expansion of our 3-D universe in more detail, Fig. 3 shows the variation of the fractional densities $F_i$ of each of its three contents. The present $R/R_0$ = 1 appears at log $R/R_0$ =0, also at redshift Z = 0 since $R/R_0 = 1/(1 + Z)$ and the curves are calculated on into the future. The symbols for the points of calculation were intentionally included to show a new feature of the SC-theory. With time defined, time t and redshift Z effectively become state variables – a great simplification in the mathematics with no integration required for the expansion. Note that the curve for dark mass $F_x$ indicates how near our universe has reached its constant rate of expansion $\dot{R}$ = c when $F_x$ = 1 at total domination by dark mass.

The universal constant $\kappa = 3/32\pi$ was used only at Z = 0 to set the present value of dark mass $\rho_{x0}$ by difference. It could have been used at every calculated point but instead, at points far from Z=0, reliance was placed on each of the expansion factors. So, as a check on the promise of the universal constant of expansion, $\kappa$ was calculated for each of the data points at $R/R_0$ in Fig. 3 as,

$$\kappa_{calc} = G[\rho_r + \rho_m + \rho_x] t^2, \tag{7}$$

and was then divided by $3/32\pi$. The horizontal line (without symbols) using the right ordinate presents the confirming results of vanishing errors of calculation with $\kappa_{calc}/\kappa$ = 1.

The modeling of the formation of the 4-D ball was presented elsewhere [12] and it did require integration of a differential equation. Fortunately an infinite series integral for a damped exponential production was available and with Planck's natural units it required only one input variable $R_F$ for control of the redshift of formation $Z_F = R_F/R_0$ of the 4-D ball. Again the use of partial times allowed a smooth connection to the equations for expansion. This connection was so fast that it all occurred between two points of Fig. 4 where time makes the vertical jump. The smooth connection between the two points will be shown later in Fig. 6. The effects of variation of $R_F$ were easily calculated [3].

In 3-D nature, things come into existence and things go out of existence. Cosmic time came into existence when the first 4-D particle was born in the epi-universe. Before that, its resistance to happening was very great. The 4-D ball and its decelerating Hubble parameter came into existence at $R/R_0 \sim 10^{-26}$ and $t \sim 10^{-34}$ s as did the mass of the elementary particles that will form matter. At formation of the 4-D ball its expansion rate was $\dot{R} \sim 10^{24}$ c which tells us immediately that the rate of transport of information in epi-space $c^+ > 10^{24}$ c, which could readily account for the non-local behavior of quantum mechanics. As per Eq. (4), tH begins at value ½ and goes to unity in the future. Note the characteristic appearance of integers and integer fractions in the limits of the development from the universal constant.

The symbol R represents the radius of the 4-D ball but it is used in Fig. 4 before the 4-D ball came into existence. Before the 4-D ball came into existence, R represents the radius of a hypothetical 4-D ball compacted of all the "free" 4-D cells then produced.

## 4. More on cosmic time and kappa

The use of partial times was explained in Section 3 to derive the universal constant of expansion. This means the rate of time changes with time during the evolution of the expansion. Can that changing rate be expressed in a single equation? First the theory says there is one length that does not change, which is the edge length $l_p$ of the 4-D hypercube. Both the epi-universe and our 3-D universe recognize that special length. The radius R of the 4-D ball increases in steps of one $l_p$ so it is the steps in time that is different between our two different universes. We must translate our normal concept of rate from the time derivative of dx/dt where the differential dt is the limit of $\Delta t \to 0$ to that of $(N\, t_p/l_p)^{-1}$ and the 3-D limiting velocity of light is $((N=1)t_p/l_p)^{-1}$.

For the expansion of the 4-D ball in the fourth dimension, epi-time is not limited to a minimum of one $t_p$ but changes can happen in fractions of one $t_p$ that is, $t_p/N$, or at rates of $((t_p/N)/l_p)^{-1}$, which can be read as $(t_p/Nl_p)^{-1}$ or many layers of $l_p$ per one $t_p$ for the high expansion rates in the early 3-D universe. Even so, nature controls the Hubble parameter to reduce $\dot{R} \to c$ or one $l_p$ per one 3-D $t_p$ or $H \to c/R \to 0$ into the future.

The differential of Eq. (3) gives dt in terms of dρ and the differential of ρ gives dρ in terms of dR. Combining differentials, and setting dR to its minimum length of $l_p$, and dividing both sides by the future time interval limit of $t_p$ gives,

$$dt/t_p = ((t_0^2/t)(\rho_0/\rho)(\rho_2/\rho)(l_p/R))/t_p. \qquad (8)$$

Equation (8) is plotted in Fig. 5 together with time t and expansion rate $\dot{R}/c$ versus log $R/R_0$.

There is another amazing feature of the universal constant of expansion. Equation (1) seems to have a simple lock on *time* that discourages its misuse. The SC-theory can be presented in the form of the Friedmann equation as an expression of the Hubble parameter squared:

$$H^2 = (\dot{R}/R)^2 = 4(G/\kappa)(\rho/(R\partial\rho/\partial R))^2 \rho. \qquad (9)$$

But to obtain the value of H required a legal mathematical operation that physically was unexpected. Taking the legal positive square root of both sides of Eq. (9) immediately exposed the lock. The partial derivative $\partial\rho/\partial R = -2\rho_2/R$ is negative which would make the Hubble parameter unphysical as negative and we know from redshift of radiation from distant stars that our universe is not collapsing. Therefore taking the positive root on the left, one must take the negative root on the right to get an acceptable H.

Before leaving this feature of the universal constant, give it one more examination. Suppose time is put back into Eq. (9) as,

$$H = ((4/t^2)(\rho/(R(-2\rho_2/R))^2)^{1/2}, \qquad (10)$$

where the positive square root on the right side is indicated but not yet performed and the perceived trouble still shows as the internal negative sign. For the imaginary time "it" of past physics, the square is $(it)^2 = -t^2$. If in Eq. (10) before obtaining the positive root, first associate the minus sign with $t^2$ to get,

$$H = (1/it)(\rho/\rho_2). \qquad (11)$$

Equation (11) is the same as Eq. (4) but in imaginary time. The universal constant of expansion seems to accept the Minkowski spacetime of the special theory of relativity with expansion, which does exist even though its effects are negligible locally.

## 5. Geometric analysis

So far the above has been a dynamic development starting with a postulated new spatial condensation dynamic and following the time-density relation t(R) of its discovered universal constant. But what of the postulated case where the 4-D ball and its 3-D surface change in size in increments of one Planck length $l_p$. Perhaps Planck's natural units will tell us more in these geometric considerations.

Some have already questioned the assumption of the continuum of space and time [13, 14], but the differential calculus will be used with lower limits of Planck's natural units. It is claimed that $l_p = 1.616 \times 10^{-33}$ cm and $t_p = 0.5391 \times 10^{-43}$ s are so very small that any mathematical errors introduced would now be far smaller than any physical measurement could detect.

The volume of the 4-D ball is $V_4 = 1/2\, \pi^2 R^4$ and that of its 3-D surface is $V_3 = 2\pi^2 R^3$. Derivatives with respect to cosmic time are,

$$\dot{V}_4 = V_3 \dot{R} = 4 V_4 H, \qquad (12)$$

$$\dot{V}_3 = 3 V_3 H. \qquad (13)$$

Assuming 3-D and 4-D cubic cells of edge length $l_p$, the number of such cells are $N_3 = V_3/l_p^3$ and $N_4 = V_4/l_p^4$ and as the universe expands, the number rates of production of these cells are,

$$\dot{N}_4 = \dot{V}_4/l_p^4 = \dot{R}\, N_3/l_p = (N_3/t_p)(\dot{R}/c), \qquad (14)$$

$$\dot{N}_3 = \dot{V}_3/l_p^3 = 3N_3 H. \tag{15}$$

In Section 3 it was shown that the present expansion rate ($\dot{R}/c$) ≈ 1 and $R_0$ is in excellent agreement with the Hubble length, $R_0 = c/H_0$, so the present values for all of the above Eqs. (12) - (15) can be computed and in particular,

$$\dot{N}_{4\,0} = (N_{3\,0}/t_p) = 1.91 \times 10^{227}\ s^{-1}. \tag{16}$$

The growing number of both types of 4-D cells, c-type $N_4$ and x-type (dark mass), are shown in the log-log plot of Fig. 6 from the first x-type cell produced to $R/R_0 = 10^3$ into the future. Although $\dot{N}_{4\,0}$ is a large magnitude for present physics, as Fig. 4 and Fig. 6 indicate, it is consistent with the postulated dynamic that drives the expansion of our universe.

It is important to show this geometric analysis agrees with the universal constant of expansion. That will be done in the limit of increasing R. From Eq. 14, $\dot{N}_4/N_3 = \dot{R}/l_p$. Equation (1) has already predicted $\dot{R} \to c = l_p/t_p$ in the limit, so $\dot{N}_4/N_3 \to 1/t_p$. A limit expression is needed from Eq. (1) for $\dot{R}/l_p$ when our universe was dominated by x-stuff in the far future:

$$\kappa = G\rho t^2 = (\rho/\rho_p)(t/t_p)^2 \to G\rho_x t^2 = G\rho_{x0} R_0^2 (t/R)^2 \tag{17}$$

or $\quad R = (G\rho_{x0}R_0^2/\kappa)^{1/2}\ t \quad$ and $\quad \dot{R} = (G\rho_{x0}R_0^2/\kappa)^{1/2}. \tag{18}$

Substituting $G = c^2 l_p/m_p$, and Planck density $\rho_p = m_p/l_p^3$, gives a new constant,

$$\dot{N}_4/N_3 = \dot{R}/l_p \to (1/t_p)[(R_0/l_p)^2(\rho_{x0}/\rho_p \kappa)]^{1/2} = 1.854 \times 10^{43}\ s^{-1} = 1/t_p, \tag{19}$$

or $\quad [(1/\kappa)(R_0/l_p)^2(\rho_{x0}/\rho_p)] = 1, \quad$ or $\quad R_0 = ct_0(\rho_0/\rho_{x0})^{1/2}, \tag{20}$

independent of R, which confirms Eq. (6). Calculation of $(\dot{N}_4/N_3)_0 = 1.863 \times 10^{43}\ s^{-1}$ but at $R/R_0 = 10^{10}$, $\dot{N}_4/N_3 = 1.855 \times 10^{43}\ s^{-1}$ in agreement with Eq. (19) Thus the geometric Eq. (14) agrees with Eq. (1) in the future limit of $\rho \to \rho_x = \rho_{x0}(R_0/R)^2$.

Recognizing that the rate of spatial condensation $\dot{N}_4$ times Planck's constant (modified to $\hbar = h/2\pi$) has units of energy and can be equated to any form of energy, therefore it is claimed that *energy* has finally been defined: *Energy, in all forms, is the rate of spatial condensation times $\hbar$,*

$$E \equiv \dot{N}_4 \hbar,\ g\ cm^2\ s^{-2}, \tag{21}$$

Applied to the present universe ($\dot{R}/c \approx 1$) this gives from Eq. (14) for the vacuum energy,

$$E_{v0} = N_{30} \hbar/t_p = 2.01 \times 10^{200}\ ergs, \tag{22}$$

The present vacuum energy density is approaching a universal fundamental constant of,

$$e_{v0} = E_v/V_3 = \hbar/l_p^3 t_p = 4.63 \times 10^{114} \text{ ergs cm}^{-3}. \tag{23}$$

The variation of these energies and densities with the expansion, as well as those of mass energy, are presented elsewhere [5]. Note that the vacuum energy density $e_{v0}$ does indeed become constant, as does Einstein's lambda.

The current ratio of vacuum energy to mass energy from Section 3 ($E_{v0}/E_{m0}$) = $2.01 \times 10^{200}/1.09 \times 10^{77} = 2 \times 10^{123}$, which is the same ratio as predicted by quantum theory if the energy sum of ground state vacuum fluctuations is cut off at the Planck level [15]. On the embarrassing impossibility of such a high vacuum **mass energy** Wilczek comments [16]:

> We do not understand the disparity. In my opinion it is the biggest and worst gap in our current understanding of the physical world.

Fortunately, the SC-theory explains *mass is not an attribute of vacuum energy*. The SC-process begins to develop *mass* for vacuum energy on the bare surface of the 4-D ball, but is immediately shut off by the new 4-D cell it just produced and so only virtual particles appear, with a momentary **trace of mass**, and then disappear.

This spatial condensation on the bare surface of the 4-D ball occurs everywhere in 3-D space and so produces 3-D space everywhere according to Eq. (13). Since it is uniform in 3-D space, so it would be in a local volume $V_3 = (4\pi/3)r^3$. Using the Gauss theorem for an imaginary 2-sphere of radius r, then $\dot{V}_3 = 4\pi r^2 v$ and solving for v gives v = Hr, which is the Hubble law for 3-D space at the Planck scale.

The effects of expansion are negligible locally, which is easily confirmed. Consider the imaginary 2-sphere around the Sun at the position of the orbit of the Earth, r ≈ 1 Au = $1.496 \times 10^{13}$ cm. With H = $2.22 \times 10^{-18}$ s$^{-1}$, the 3-D space flowing through the Earth is only $v_s = Hr = 4.98 \times 10^{-5}$ cm/s. Even at the radius of our orbit around the Galaxy r ≈ 8 kpc, the 3-D velocity $v_s = Hr = 55 \times 10^3$ cm/s. Changing origins, the conventional peculiar velocity $v_p = -v_s$ is that of the mass through space. Not until radial distances ~ 1 Mpc, can the present drag $a_r = -Hv_p$ of 3-D space nullify the negative acceleration due to gravity of large galactic masses [17].

The fundamental unit of *action*, $\hbar$, has dimensions of *energy* times *time*. The variational principles on this entity are very powerful [18]. But like many other fundamental entities in physics, the mathematical rules for its use have been developed without any knowledge of why they work. In the SC-theory a unit of *action* is the production of one 4-D cell in one Planck time and all such cells produced since the beginning of cosmic time are stored in the growing 4-D ball. This suggests that the important concept of "least action" translates to a frugal epi-space that tends to guide our 3-D motions towards trajectories with the "least spatial condensation," $N_4 \hbar = \dot{N}_4 \hbar \delta t$.

There is much more from the SC-theory, but this short geometric unfolding at the present time does indeed support the full SC-theory of Section 3. For an example of an SC-prediction far from the present, the reader may have wondered how the SC-theory performs in the early era of nucleosynthesis. Carroll and Kaplinghat published a recent paper [19] and their Fig. 6 showed acceptable model curves of log(H) versus log(a/a$_0$) during nucleosynthesis. The SC-curve is parallel to the standard model curve (slightly higher) and meets their minimum range of H at $a/a_0 = 8.5 \times 10^{-10}$ with $T_e = 0.277$ Mev.

## 6. Problems of present big bang theory

This paper will end with a brief attempt to explain the global problems of present theory, which are by no means minor as described by theoretician T. Padmanabhan on the addition of "dark energy" to our universe,

> …The cosmological observations have thrust upon us a rather preposterous composition for the universe which defies any simple explanation, thereby posing the greatest challenge theoretical physics has ever faced.
>
> Current Science, paper 1057 by T. Padmanabhan, *Dark Energy: The cosmological challenge of the century* in the Special Section: Cosmology and arXiv.astro-ph/0411044. Used with permissions.

Some scientists now claim that observations tend to rule out all present cosmologies [20] and that we may have to abandon the Friedmann equation [21].

The source of the GR-problem is the global misunderstanding of the fundamental role of time in the evolution of our universe. Einstein relegated time to the fourth geometric dimension of our universe. That role was adequate for local physics and its symmetric parametric (clock) time. But as we have seen in the SC-theory, global time represents something much more complex than the block time of his geometry.

That local role for time in general relativity meant that the Friedmann equations were forced to place the major role for the expansion in terms of the Hubble parameter squared. Also note that there is a natural Planck time but not a natural Planck Hubble parameter.

With only three spatial dimensions, the three GR-singularities (beginning, closed end, and inside a black hole) warned of trouble ahead as did the early particle horizon. The added very early period of inflation was only a patch, but the recent addition of "dark energy" mass density forced the expansion to accelerate. It is the effect of this added acceleration on the interplay of time and the Hubble parameter that is now addressed.

Peacock [6], presents his GR-equation for t(R) [Eq. 3.44], (here, for k = 0) as,

$$t(Z) = H(Z)^{-1}(1 - \Omega(Z))^{-1}. \tag{24}$$

Since for matter in the future, $\Omega(Z)$ goes quickly to zero, this equation is essentially the same as the SC-Eq. (4) of Section 3 that says for the present and into the future, $tH \approx 1$ or t and H change inversely into the future. Also Peacock states [6] that $\Sigma\Omega_i = 1$ at all times for a flat universe of curvature k = 0.

In terms of the Hubble parameter H(Z), the equation of Davis and Lineweaver for the BB-model is [22],

$$H(Z) = H_0(1+Z)[1 + \Omega_m Z + \Omega_r((1+Z)^2-1) + \Omega_\Lambda((1+Z)^{-2} - 1)]^{1/2}, \tag{25}$$

Substituting $R = R_0/(1 + Z)$ into Eq. (25) and dividing by c gives,

$$\dot{R}/c = (R_0 H_0/c)[1 + \Omega_m Z + \Omega_r((1+Z)^2-1) + \Omega_\Lambda((1+Z)^{-2} - 1)]^{1/2}. \tag{26}$$

Using $\Omega_\Lambda = 1 - \Omega_r - \Omega_m$, drives GR - $\dot{R}/c$, and thus GR-H(Z) unphysical towards infinity in the future as shown in Fig. 7 in contrast to the SC-predicted curve of H(Z) going to zero, according to Eq. (A.16).

The grossly unphysical prediction of the BB-theory, shown in Fig. 8, is that GR-time is driven to zero in the future by the acceleration inversely to the very rapid rise of the Hubble parameter. In contrast, the SC-time rises rapidly into the future, as one would expect from Eq. (A.1), or $t = ((\kappa/G)/\rho_T)^{1/2} \to \infty$, as $\rho_T \to 0$.

Note that in units of Z, the entire future of our 3-D universe is contained between $Z = 0$ and $Z = -1$; ($R/R_0 = 1/(1 + Z)$). If present physicists had carried their calculations on into the future, that step would have disclosed other unphysical predictions.

## 7. Summary and conclusions

Once the universal constant of expansion was deciphered, the SC-vision and it told an inspiring story of how our universe was formed and how it expands today using only simple mathematics. It did not show that Einstein's predictions of local physics were wrong but suggested his theory applied more to the physics of the older interacting epi-universe. Only at one point does his local physics fail and and that is at the singularity inside a black hole. Even there the Schwarzschild metric makes a noble effort with a distorted 3-D hyperspace, but *time* as the fourth dimension cannot relieve a 3-D infinite density of mass.

The SC-theory became a tight theory with the replacement of the unknown dark matter of present physics with the new SC-4-D x-stuff. That x-stuff dark mass scales differently than matter and set the expansion of our universe on the correct evolutionary path according to the new universal constant kappa.

The proposed change in topology of replacing Einstein's fourth dimension of time with a fourth dimension of space allowed the new dynamic of spatial condensation to do its work and account for many things. These included a beginning of our 3-D universe without a singularity and an explanation of gravity by spatial condensation on masses that locally curve 3-D space without the need of the mysterious concept of an attractive force.

The concept of time required considerable revision including the new feature that cosmic time itself changes rate as energy domination of the universe changes slowly from radiation to matter and then to dark mass.

Finally we come to the feature of the new theory that may dominate theoretical study in the future. That is the postulated older epi-universe that spawned our 3-D universe. There is much of our present physics that may only be accounted for by the interactions of the older, very different epi-universe. That prediction and the new dynamic of spatial condensation may hold the key to the future unification of physics.

The paper to follow will undertake the task of explaining and predicting the transport of radiation on the surface of the 4-D ball. The immediate goal will be to predict the measurements of supernova Ia radiation from these exploding stars with the new SC-theory that does not permit acceleration of its expansion rate.

## Acknowledgement

The author thanks his good friend, Emeritus Professor Robert A. Piccirelli, for editorial comments

**Reference**s
1. Ellis, George F. R. Issues in the Philosophy of Cosmology. http://xxx.arxiv.org/abs/astro-ph/0602280)(2006).
2. Sciama, D. W. *Modern Cosmology.* (Cambridge University Press, Cambridge) (1971).
3. Leffert, C. B, *Evolution of Our Universe: via Spatial Condensation* (Anoka Publishing, Troy) (1999).
4. Leffert, C. B. Supernova Ia Predicted without the Cosmological Constant. (http://xxx.arxiv.org/abs/astro-ph/0102071) (2001).
5. Leffert, C. B. Supernova Ia without Accelerated Expansion: The First Global Failure of Relativity Theory. (http://xxx.arxiv.org/abs/astro-ph/0501176) (2005).
6. Peacock, J. A. *Cosmological Physics.* (Cambridge University Press, Cambridge), (1999). p 75, 78
7. Whitrow. G. J. *The Natural Philosophy of Time.* (Thomas Nelson and Sons, New York) (1961) p 253
8. Wagoner, R. V. ApJ, **179**, (1973). p. 343
9. Kolb, E. W. & Turner, M. S. *The Early Universe.* (Addison-Wesley, Reading) (1990) p 98 & p 84.
10. Schwarzschild, B. *Physics Today* June 17 –19, (1998).
11. Spergel, D. N. et al, (http://xxx.arxiv.org/abs/astro-ph/0302209) (2003).
12. Leffert, C. B. 1995 *Time and Cosmology: Creation and Expansion of Our Universe.* (Anoka Publishing, Troy) (1995).
13. Penrose, R. "On the Nature of Quantum Geometry" in *Magic without Magic: John Archibald Wheeler*, Ed. J. R. Klauder, (W. H. Freeman & Co., San Francisco) (1972).
14. Wigner, E. P."Response" in *New Techniques and Ideas in Quantum Measurement Theory*. Ed. D. M. Greenberger (New York Academy of Sciences, New York) (1986). p xix.
15. Sahni, V. (http://xxx.lanl.gov/abs/astro-ph/0202076) (2002).
16. Wilszek, F. (http://xxx.lanl.gov/abs/hep-ph/0101187) (2001). p. 23.
17. Leffert, C. B. *Large-Scale Structure from Spatial Condensation and Reproducing Dark Mass.* (http://xxx.arxiv.org./abs/astro-ph/0106236) (2001).
18. Lanczos, C. *The Variational Principles of Mechanics.* (Dover Publications. Inc., New York) (1970).
19. Carroll, S. M. & Kaplinghat, M. Phys. Rev D65 063507 (http://xxx.arxiv.org/abs/astro-ph/0108002).(2001)
20. Vishwakarma, R. G. (http://xxx.arxiv.org/astro-ph/0511628).(2005).
21. Shapiro, C. & Turner, M. S. (http://xxx.arxiv.org/abs/astro-ph/0512586) (2005).
22. Davis, T. M. & Lineweaver, C. H. (http://xxx.arxiv.org/abs/astro-ph/0310808) (2003)

# Table A: Summary of the SC-Cosmological Theory

The scale factor R has units of length for our 3-sphere, spatially 3-dimensional expanding universe; G is the gravitational constant; c is the local speed of light; and H is the Hubble parameter. Present values have subscript 0 and cgs units are assumed. Other subscripts include: r=radiation, m=matter and x=dark mass (not dark matter). Pertinent equations of the new theory [hereafter: "SC-theory"] are listed in Table A1.

### Table A.1  Derivation of Theory

| | | |
|---|---|---|
| Universal constant: | $\kappa = Gt^2\rho = Gt_0^2\rho_0 = 3/32\pi$. | (A.1) |
| From $T_0$=2.726 K: | $\rho_{r0} = 9.40 \times 10^{-34}$ g cm$^{-3}$. | (A.2) |
| From nucleosynthesis: | $\rho_{m0} = 2.72 \times 10^{-31}$ g cm$^{-3}$. | (A.3) |
| Present age: | $t_0 = 13.5$ Gy $= 4.260 \times 10^{17}$ s. | (A.4) |
| From (A.1): | $\rho_0 = (\kappa/G)/t_0^2$ g cm$^{-3}$. | (A.5) |
| From (A.6): | $\rho_{x0} = \rho_0 - \rho_{r0} - \rho_{m0}$ g cm$^{-3}$. | (A.6) |
| Present scale factor: | $R_0 = ct_0(\rho/\rho_{x0})^{1/2}$ cm. | (A.7) |
| Redshift Z (Input): | $R = R_0/(1+Z)$ cm. | (A.8) |
| Radiation density: | $\rho_r = \rho_{r0}(R_0/R)^4 = \rho_{r0}(1+Z)^4$. | (A.9) |
| Matter density: | $\rho_m = \rho_{m0}(R_0/R)^3 = \rho_{m0}(1+Z)^3$. | (A.10) |
| Dark Mass density: | $\rho_x = \rho_{x0}(R_0/R)^2 = \rho_{x0}(1+Z)^2$. | (A.11) |
| Total density: | $\rho(R) = \rho_r + \rho_m + \rho_x$. | (A.12) |
| Cosmic time: | $t(R) = + (t_0^2 \rho_0/\rho(R))^{1/2}$. | (A.13) |
| From time derivative: | $\rho_2 = 2\rho_r + 3/2\,\rho_m + \rho_x$. | (A.14) |
| From time derivative: | $\rho_3 = 4\rho_r + 9/4\,\rho_m + \rho_x$. | (A.15) |
| From time derivative: | $H = \dot{R}/R = (\rho/\rho_2)/t$. | (A.16) |
| Expansion Rate: | $\dot{R}/c = (R/ct)(\rho/\rho_2)$. | (A.17) |
| Deceleration, $q = -\ddot{R}R/\dot{R}^2 =$ | $(1/Ht)(-1 + [3 - 2(\rho\rho_3/\rho_2^2)])$. | (A.18) |

The scaling with the expansion of radiation, Eq. (A.9), and matter, Eq. (A.10), are borrowed from the big bang model, $\kappa$ has the same value for early Friedmann radiation.

The postulated scaling, Eq. (A.11), of the new and now dominant stuff called "dark mass," is the key signature of this new cosmological theory. Its density decreases with the expansion but its total mass, always in individual clumps, increases with the expansion. It is not a 3-D substance and so does not interact with radiation or matter except gravitationally, where it certainly contributes to the local curvature of 3-D space. The distribution of these miniscule dark mass seeds at the beginning of the expansion sets the pattern for the present large-scale structure, including voids, and contributes to the early formation of black holes and fit to supernova Ia data for $t_0$=13.5 Gy with no acceleration of the expansion rate.

The basic postulate for cosmic time, Eq. (A.13), was made in terms of partial times $\Gamma_i$ where $t^{-2} = \sum_i \Gamma_i^{-2}$ and $\Gamma_i^2 = (\kappa/G)/\rho_i(Z)$ where $\rho_i$ are given by Eqs. (A.9) to (A.11). With age set to $t_0$=13.5 Gy, the SC-theory predicted the following values for the present cosmological parameters: $R_0$=1.354x10$^{28}$ cm, $H_0$=68.6 km s$^{-1}$ Mpc$^{-1}$, $\Omega_B$=0.031, $\Omega_{DM}$=0.248, $\Omega_{DM}/\Omega_B$=8.0, $(\dot{R}/C)$=1.005 and $q_0$=0.0084 (i.e., approaching steady-state expansion), all within the range of uncertainty of our astronomer's measurements.

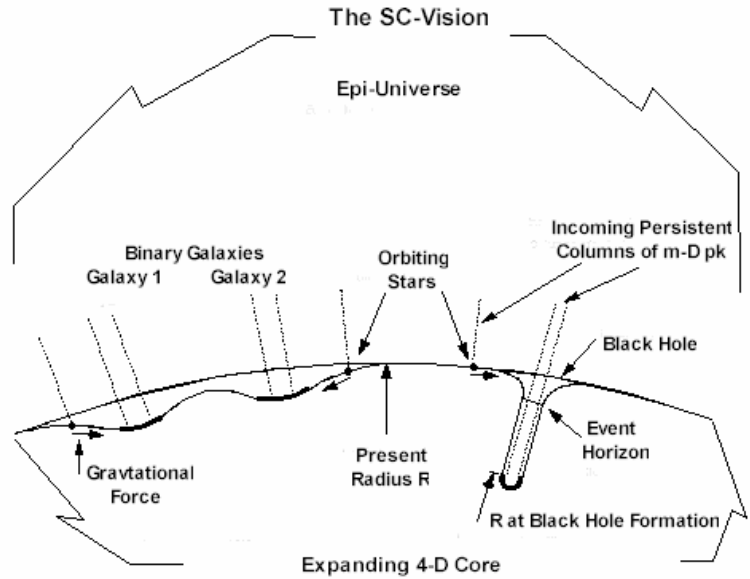

Fig. 1 Spatial condensation on the contents of the surface of the 4-D ball is shown for matter and dark mass (mass-energy) by incoming persistent columns of m-D cells. Fluctuating spatial condensation (vacuum energy) to the bare surface of the 4-D ball can be probed by Casimir-type experiments.

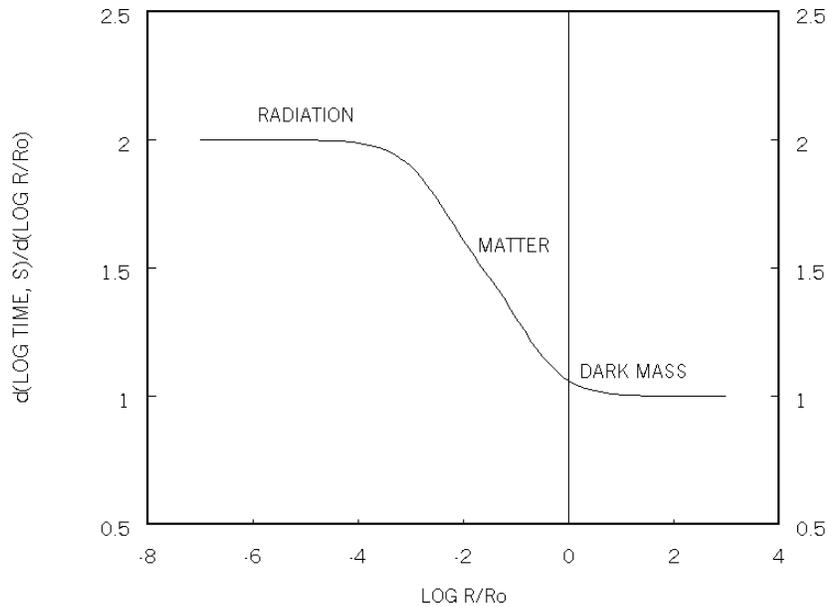

Fig. 2 Time is unique in the SC-theory in that it is defined in terms of the changing resistance to spatial condensation as the dominant mass energy evolves with the expansion from radiation to matter and finally to dark mass.

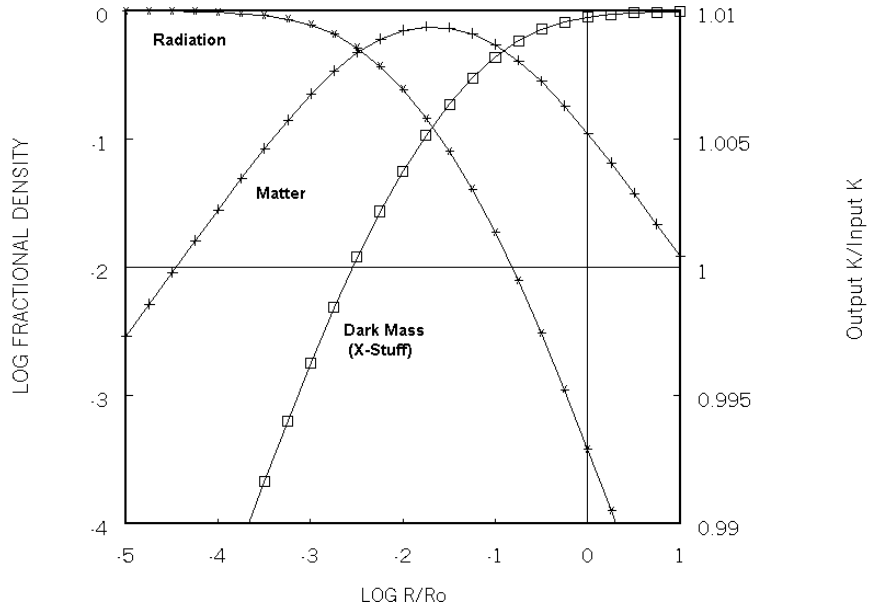

Fig. 3 Evolution of the three contents of our 3-D universe is shown in terms of log of fractional densities $\rho_i/\rho$ versus the log of its size $R/R_0$. For the SC-theory, the decoupling of matter and radiation at temperature $T \approx 0.26$ ev occurs at log $R/R_0 \approx -3.0$ before the equality of matter and radiation at log $R/R_0 \approx -2.5$.

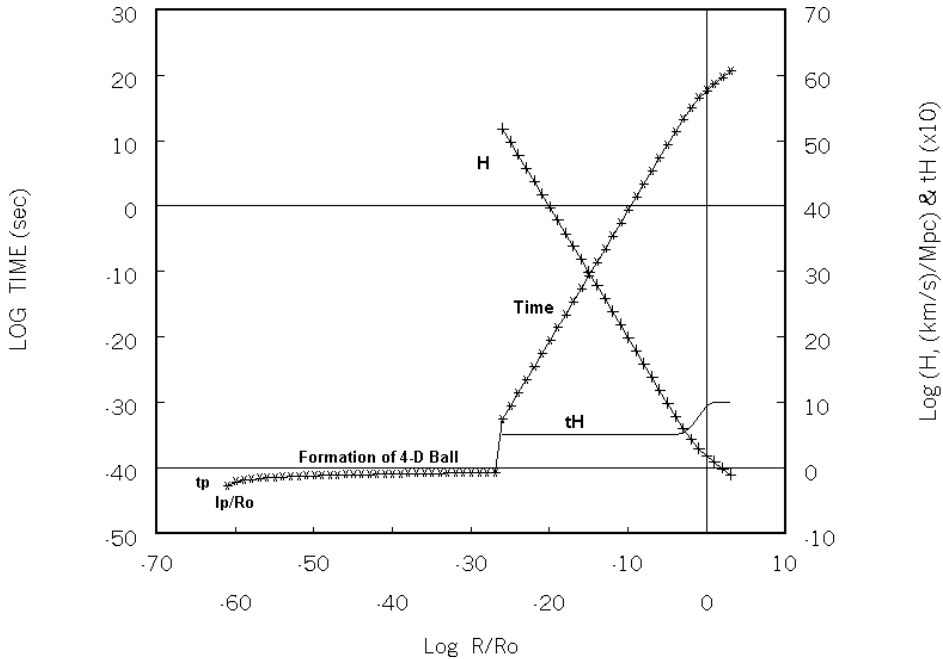

Fig. 4 The compaction of the 4-D ball and its further expansion are joined with cosmic time which starts at Planck time $t_p$ producing free 4-D cells, which are then compacted during the rapid rise of time. The Hubble parameter comes into existence with the completion of the 4-D ball whereupon time and H change reciprocally as in their product shown as the tH curve without points.

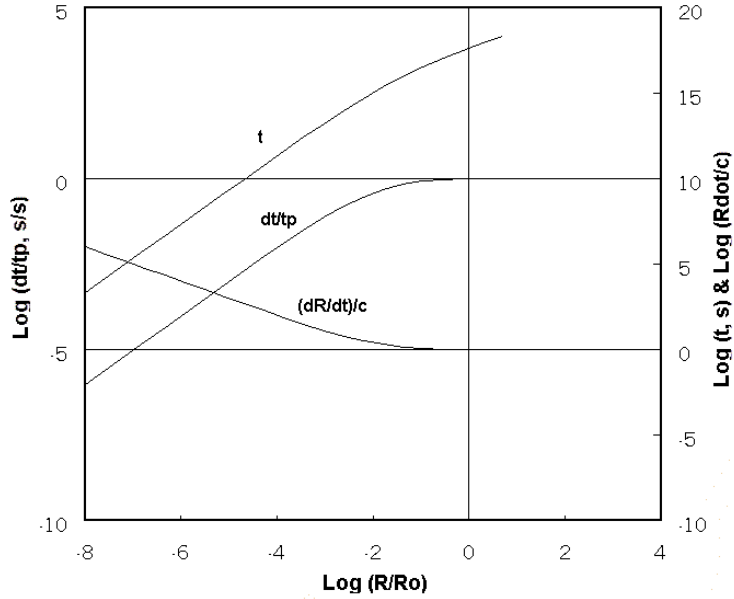

Fig. 5 The dimensionless expansion rate $\dot{R}/c$ of our universe has now almost reached its limiting value of unity or R = ct, in agreement with the fourth dimension of Einstein's block spacetime. The rate of SC-cosmic time, relative to the Planck interval of time, $dt/t_p$ changes with time and approaches unity into the future.

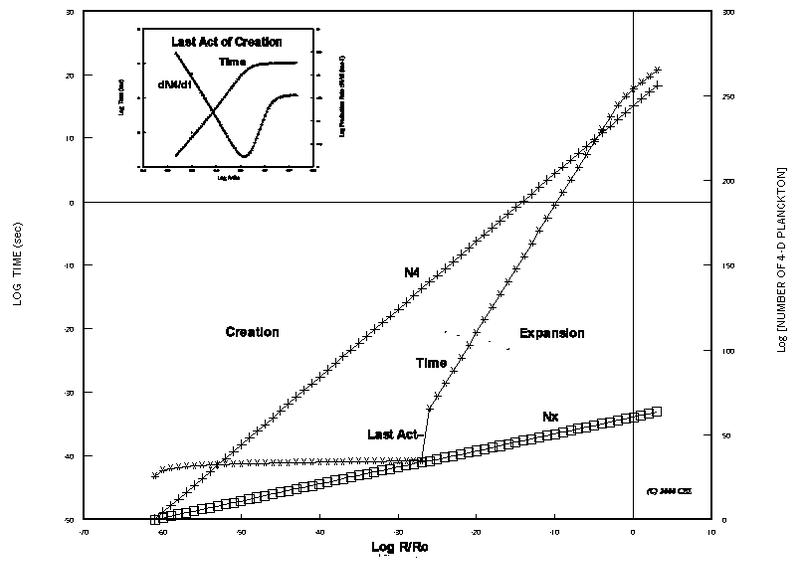

Fig. 6 Combining the beginning with the expansion of our 3-D universe shows the smooth production of both c-type ball-acceptable 4-D cells $N_4$, and x-type 4-D cells $N_x$ of dark mass. With 64 decades of size on the abscissa, the rapid rise of time at log $R/R_0$ from –27 to –26 misses all of the detail, partly shown in the overlay for log $R/R_0$ from –26.24 to –26.19. The beginning starts at $Z=10^{61}$, the expansion at $Z=10^{26}$, and the present at Z=0.

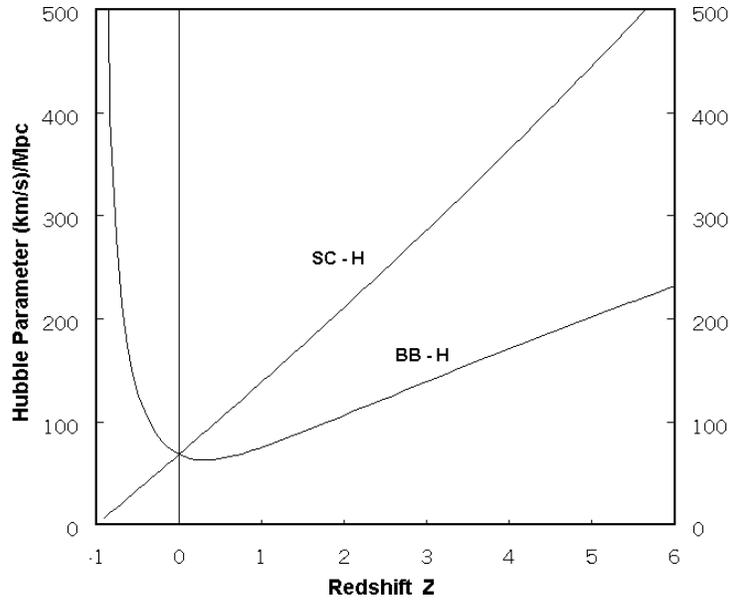

Fig. 7 In the SC-theory all densities go to zero in the far future as does the Hubble parameter, SC-H. The big bang H was also headed toward zero in the future but the added energy of the cosmological constant to make $\Omega = 1$ creates the acceleration of the unphysical drive of BB-H to infinity in the far future.

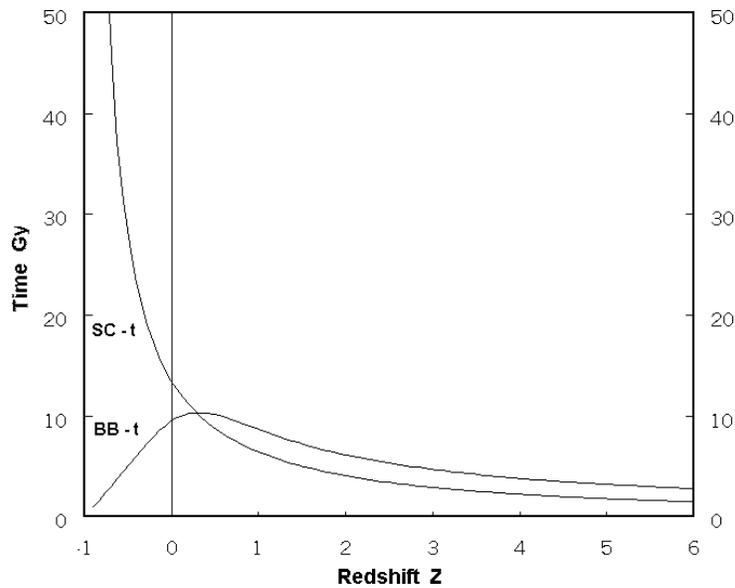

Fig. 8 Both SC-theory and BB-theory predict tH goes to unity in the far future. Thus time is driven in the opposite direction to that of H in Fig. 7. Here SC-time increases without limit as the universe expands as we expect, but the added acceleration to the BB-model reverses BB-time with an unphysical drive to zero in the far future.